\documentclass{procpavia}
\newcommand{\beq} {\begin{equation}}
\newcommand{\eeq} {\end{equation}}

\newcommand{\pon} {p_1}
\newcommand{\ptw} {p_2}

\newcommand{\bpi}{{\bf p}}
\newcommand{\bqu}{{\bf q}}

\newcommand{\half}{\frac{1}{2}}
\begin{document}
\pagestyle{prochead}


\title{ TWO PROTON EMISSION WITH ELECTROMAGNETIC PROBES}
\author{A.M. Lallena} \email{lallena@ugr.es} \affiliation
{Departamento de F\'{\i}sica Moderna, Universidad de Granada, E-18071
Granada, Spain.}  \author{M. Anguiano}
\email{marta.anguiano@iec.csic.es} \affiliation {Departamento de
Radiaci\'on Electromagn\'etica, Instituto de F\'{\i}sica Aplicada,
CSIC, \\c/ Serrano 144, E-28006 Madrid, Spain.}  \author{G. Co'}
\email{Giampaolo.Co@le.infn.it} \affiliation {Dipartimento di Fisica
Unversit\`a di Lecce and I.N.F.N 
Sez. di Lecce, \\ I-73100 Lecce, Italy\\~\\}

\begin{abstract}
A model to study two-proton emission from nuclei induced by
electromagnetic probes is developed. The process is due to one-body
electromagnetic operators, acting together with short-range
correlations, and two-body $\Delta$ currents. The model includes
all the diagrams containing a single correlation function. The
sensitivity of the cross section to the details of the correlation
function is studied by using realistic and schematic correlations.
Results for the $^{16}$O nucleus are presented.
\end{abstract}
\maketitle
\setcounter{page}{1}


\section{Introduction}

In this work we apply the model we have developed in these last years
\cite{mok00}-\cite{ang02} to describe electromagnetic responses by
considering also short-range correlations (SRC) to the case of
the electromagnetically induced two-nucleon knockout from nuclei. 
The aim of our model is to take into account the 
SRC in electromagnetic processes involving nuclei with $A>4$, in
such a way that different processes, like inclusive electron
scattering, one- and two-nucleon emission induced by electron
scattering and, eventually, real photon scattering, can be described
by using the same methodology. The idea was to have a consistent view
of the different processes.

The linear response of the nucleus 
to an external operator $O(\bqu)$ can be written as:
\[ S(\bqu,\omega) \, = \, 
- \frac{1}{\pi} \, \rm{Im} \, D(\bqu,\omega) \, ,
\]
with
\[
D(\bqu ,\omega) \, = \, \sum_n \, \xi_n^+(\bqu)
\, (E_n - E_0 - \omega + i \eta)^{-1} \, \xi_n(\bqu) 
\]
where we have defined 
\[
\xi_n(\bqu) \, = \, \frac { \langle \Psi_n | \,O(\bqu)\, |\Psi_0 \rangle
 } {\langle \Psi_n |\Psi_n\rangle ^{\half}\,\langle \Psi_0
 |\Psi_0\rangle ^{\half} }
\]
This function involves the transition matrix element between the
initial and the final states of the nucleus.
These states are construct by acting with
a correlation operator $F$ on uncorrelated Slater determinants, as
established by the Correlated Basis Function theory:
\[
|\Psi_0\rangle \,= \,F \,|\Phi_0\rangle \,\,\,\,\,\, 
|\Psi_n\rangle \,= \,F \,|\Phi_n\rangle \, .
\]
Then the quantity $\xi$ can be written as 
\[
\xi_n(\bqu) \, = \, \frac {\langle \Phi_n |\,F^+ \,O(\bqu) \,F|\Phi_0 \rangle
}{\langle \Phi_0|\,F^+\,F\,|\Phi_0\rangle } \left[ \frac {\langle \Phi_0
|\,F^+\,F\,|\Phi_0\rangle }{\langle \Phi_n|\,F^+\,F\,|\Phi_n\rangle }
\right]^\half \, .
\]

We consider scalar correlation functions of the type: 
\[
F \,=\, \displaystyle \prod_{i<j}\, f_{ij} \, ,
\]
therefore the $\xi$ function results
\[
\xi_n(\bqu) \, = \, \frac {\langle
\Phi_n |\,O(\bqu)\, \displaystyle 
\prod_{i<j}\,(1+h_{ij}) \,|\Phi_0 \rangle } {\langle
\Phi_0| \displaystyle \prod_{i<j}\,(1+h_{ij}) \,|\Phi_0\rangle }\, 
\left[ \frac {\langle
\Phi_0 | \displaystyle \prod_{i<j}\,(1+h_{ij}) \,|\Phi_0\rangle }
{\langle \Phi_n| \displaystyle \prod_{i<j}\,(1+h_{ij}) \, |\Phi_n\rangle } 
\right]^\half \, , 
\]
where the the $h$ function is defined as:
\[ h_{ij}\, =\, f^2_{ij}-1 \, .
\]

We perform the full cluster
expansion of this expression and this allows us to eliminate the
unlinked diagrams. At this point we insert the main approximation of
our model by truncating resulting expansion such as
only those diagrams which involve a single correlation
function $h$ are retained:
\[
\xi_n(\bqu) \rightarrow \xi^1_n(\bqu) \,=\,
\langle \Phi_n |\,O(\bqu)\, \left( 1+\sum_{i<j}\, h_{ij} \right) 
\,|\Phi_0 \rangle _L \, .
\]
In the above expression 
the subscript $L$ indicates that only linked diagrams are
included in the expansion. 

For one-particle one-hole final states, the function
$\xi$ includes three terms:
\[
\xi^1_{\rm 1p1h}(\bqu) \, = \, 
\langle \Phi_{\rm 1p1h} |\, O(\bqu)\, |\Phi_0 \rangle
+ \, \langle \Phi_{\rm 1p1h} |\, O(\bqu) \, 
\displaystyle
\sum^{A}_{j>1}\, h_{1j}\, |\Phi_0 \rangle
+ \, \langle \Phi_{\rm 1p1h} |\, O(\bqu) \, 
\displaystyle
\sum^{A}_{1< i <j } \, h_{ij} \, |\Phi_0 \rangle
\]

\begin{figure}[b]
\begin{center}
\includegraphics[height=0.3\textheight]{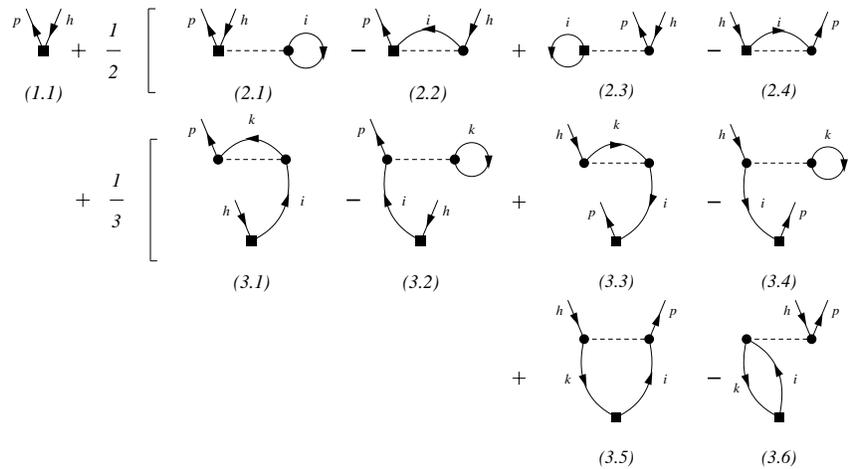}
\caption{\label{fig:diag-1p1h} 
\small Meyer-like diagrams contributing to $\xi^1_{\rm 1p1h}(\bqu)$.}
\end{center}
\end{figure}

\begin{figure}[htb]
\begin{center}
\includegraphics[height=0.3\textheight]{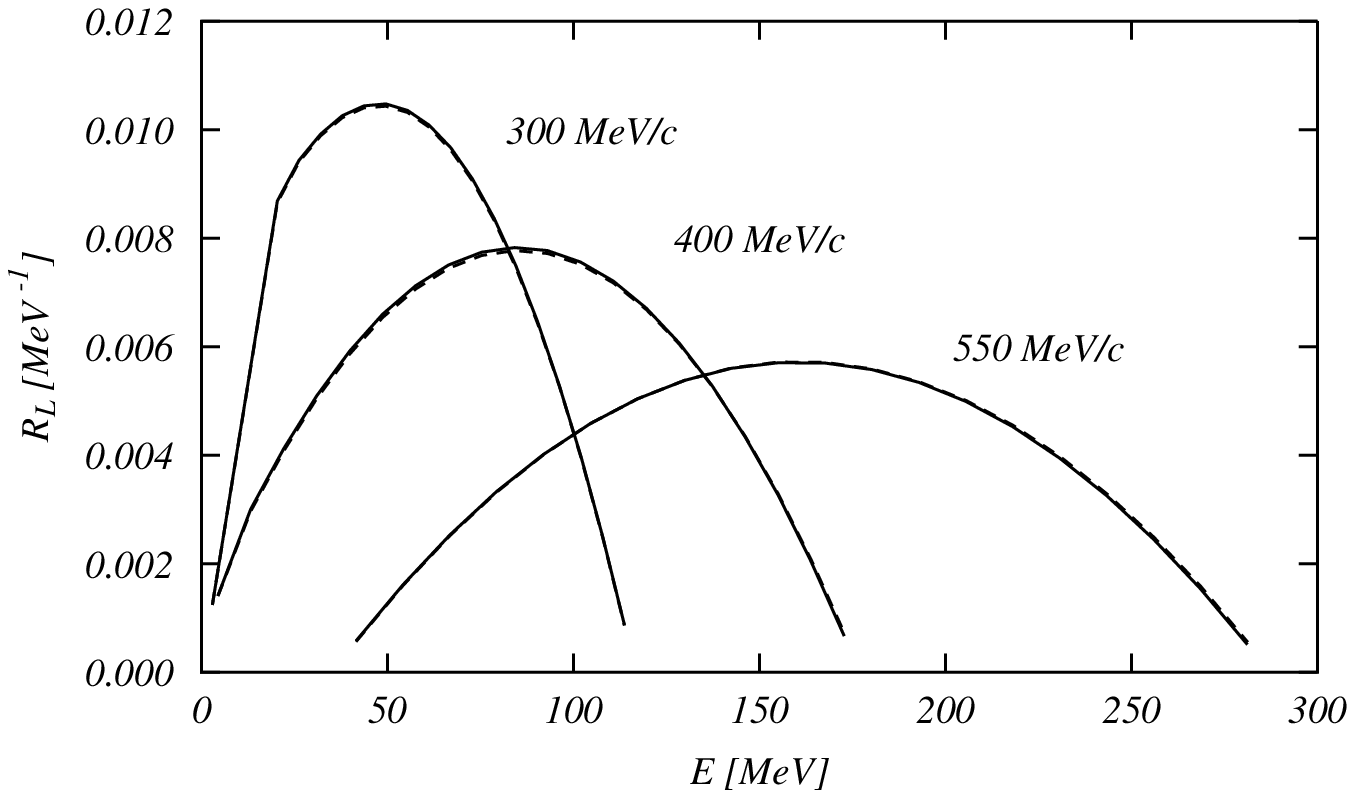}
\caption{\label{fig:1p1h-comp} 
\small Nuclear matter one-particle one-hole longitudinal response
calculated with our model (full lines) and with FHNC (dashed lines)}
\end{center}
\end{figure}

The contributions of the various terms of the above expression are
shown in Fig. \ref{fig:diag-1p1h} in terms Meyer-like diagrams.
In addition to the uncorrelated
transition represented by the one-point diagram (1.1). But, besides,
also 4 two-point diagrams and 6 three-point diagrams are
present. These new terms are necessary to get a proper normalization
of the nuclear wave functions. In the figure, the black squares
represent the points where the external operator is acting, the dashed lines
represent the correlation function $h$ and the continuous oriented
lines represent the single-particle wave functions. The letters $h$,
$i$ and $k$ label holes, while $p$ labels a particle. A sum over $i$
and $k$ is understood.

Our model was tested by comparing our nuclear matter charge response
functions with those obtained by considering the full cluster
expansion \cite{ama98}. As we can see in Fig. \ref{fig:1p1h-comp}
both calculations overlap and this gave us confidence to
extend the model to other situations.

With this model we have analyzed different processes involving both
real and virtual photons. Inclusive (e,e') excitation of discrete
nuclear states showed a very small effect due to SRC \cite{mok00}. On
the other hand, inclusive electron scattering in the quasi-elastic
peak region presented SRC contributions smaller than final state
interactions effects \cite{co01}. These last also dominate in the case
of (e,e'p) reactions \cite{mok01}. Finally, in ($\gamma$,p) we found a
kinematic region showing pronounced sensitivity to the
SRC. Unfortunately, in this region, the effect of 
meson exchange currents (MEC) is much larger than that of SRC
\cite{ang02}.

\begin{figure}[htb]
\begin{center}
\includegraphics[height=0.4\textheight]{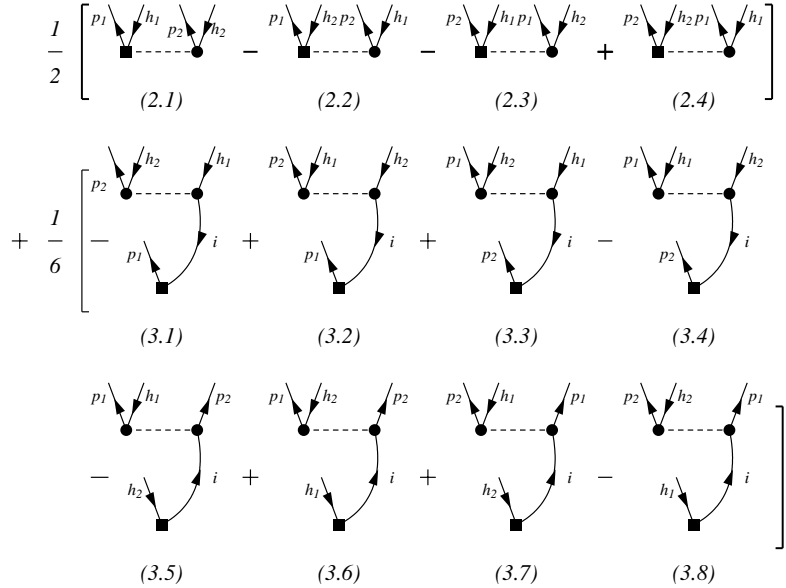}
\caption{\label{fig:diag-2p2h} 
\small Meyer-like diagrams contributing to $\xi^1_{\rm 2p2h}(\bqu)$.}
\end{center}
\end{figure}

\begin{figure}[htb]
\begin{center}
\includegraphics[height=0.3\textheight]{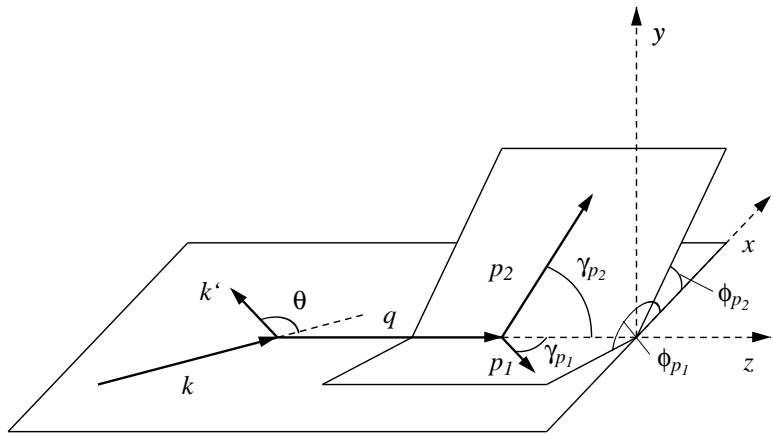}
\caption{\label{fig:geometry} 
\small Geometry of the (e,e'2p) process.}
\end{center}
\end{figure}

\begin{figure}[htb]
\begin{center}
\includegraphics[height=0.28\textheight]{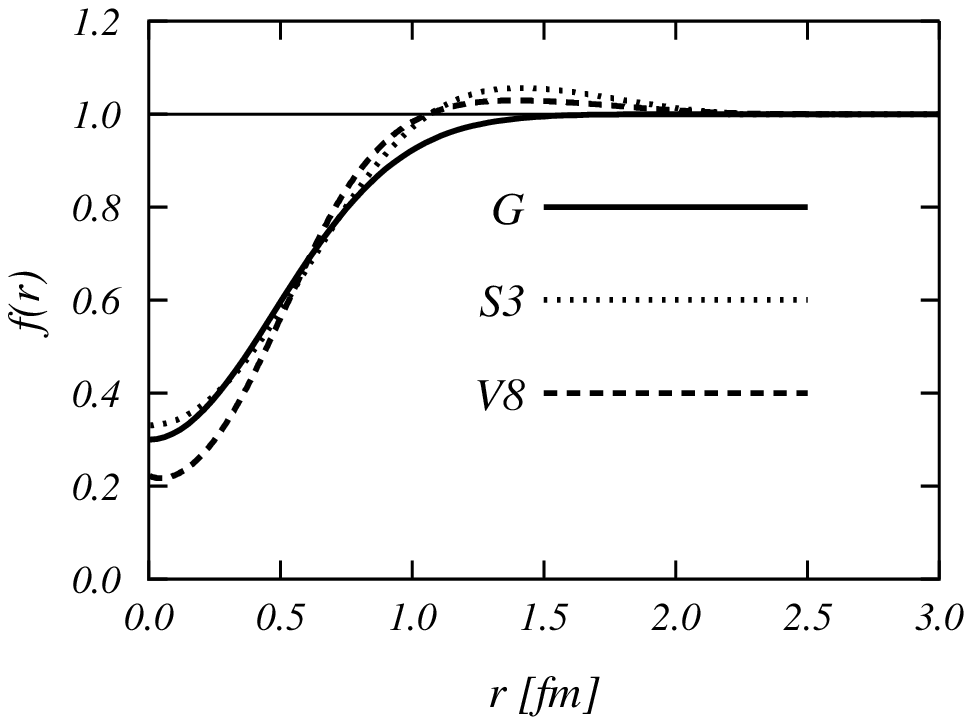}
\caption{\label{fig:corr} 
\small Correlation functions used in our calculations.}
\end{center}
\end{figure}

In all these processes the uncorrelated one-body response dominates
the cross sections. To get rid of this contribution one has to
investigate two nucleon emission processes.
In this case one
deals with two-body operators which can be either one-body operators
acting in correlated nuclear states or MEC. If, in addition, one
chooses two-proton emission, MEC involving charged mesons do not
contribute and the situation is, at least in principle, rather clean
to investigate SRC effects.

In case one has two-particle two-hole final states, 
the function $\xi$ includes two terms: 
\[
\xi^1_{\rm 2p2h}(\bqu) \, = \, \langle \Phi_{\rm 2p2h} |\,O(\bqu) \,
\displaystyle
\sum^{A}_{1<j}\,h_{1j}\, |\Phi_0 \rangle \, + \, 
\langle \Phi_{\rm 2p2h} |O(\,\bqu) \,
\displaystyle
\sum^{A}_{1< i  <j } \,h_{ij}\, |\Phi_0 \rangle
\]
The first term (see Fig. \ref{fig:diag-2p2h}) consists of 4
two-point diagrams. The second one results from the sum of 8 
three-point diagrams. As in the previous case,
this set of diagrams conserves the correct normalization.

The geometry of the two-proton emission process produced by electron
scattering from nuclei is shown in Fig. \ref{fig:geometry}. The cross
section is given by:
\[
\frac{{\rm d}^8 \sigma}
{{\rm d}\epsilon' {\rm d}\Omega_e 
 {\rm d}\epsilon_1 {\rm d}\Omega_{\pon} {\rm d}\Omega_{\ptw} } 
\, = \, \frac{m^2 \, |\bpi_1| \, |\bpi_2|}{(2 \pi)^6} \, 
\sigma_{\rm Mott} \, f_{\rm rec} 
\left(  v_l w_l + v_t w_t + v_{tl} w_{tl} 
             + v_{tt} w_{tt}   \right) \, ,
\]
where $\epsilon'$ is the energy of the outgoing electron, $\Omega_e$
is the solid angle of its direction, $\epsilon_1$ is the energy of the
one emitted proton, $\Omega_{\pon}$ and $\Omega_{\ptw}$ are the
solid angles of the directions of the two emitted protons, $\bpi_1$
and $\bpi_2$ are their momenta, $m$ is the proton mass, $\sigma_{\rm
Mott}$ is the Mott cross section and $f_{\rm rec}$ describes the
recoil of the residual nucleus. The $v$'s are kinematic factors,
while the $w$ responses include the information about the nuclear
structure and are given in terms of matrix elements of the charge and
current operators.

In our calculations we have considered 
the charge and the magnetization current one-body operators
and the $\Delta$-isobar two-body current, specifically these terms
proportional to $\tau_3$ \cite{ang03}. 

The hole wave functions have been obtained from a Woods-Saxon
potential whose parameters have been fixed to reproduce rms charge
radii and single particle energies around the Fermi energy, while
for the particles wave functions we used the optical potential
of Schwandt {\it et al.} \cite{sch82}.

\section{Results}

The particular aspect we have investigated concerns to the
sensitivity of the (e,e'2p) cross section to the details of the
SRC. This has been done by using the three correlation functions shown
in Fig. \ref{fig:corr}. The G (Gaussian) and S3 correlations, have
been taken from a FHNC calculation done with
semi-realistic interaction \cite{ari96}.  The V8 correlation is the
scalar part of a state dependent correlation used in FHNC calculation
done with a V8' Argonne interaction plus three-body Urbana IX
interaction \cite{fab00}.

\begin{figure}[htb]
\begin{center}
\includegraphics[height=0.70\textheight]{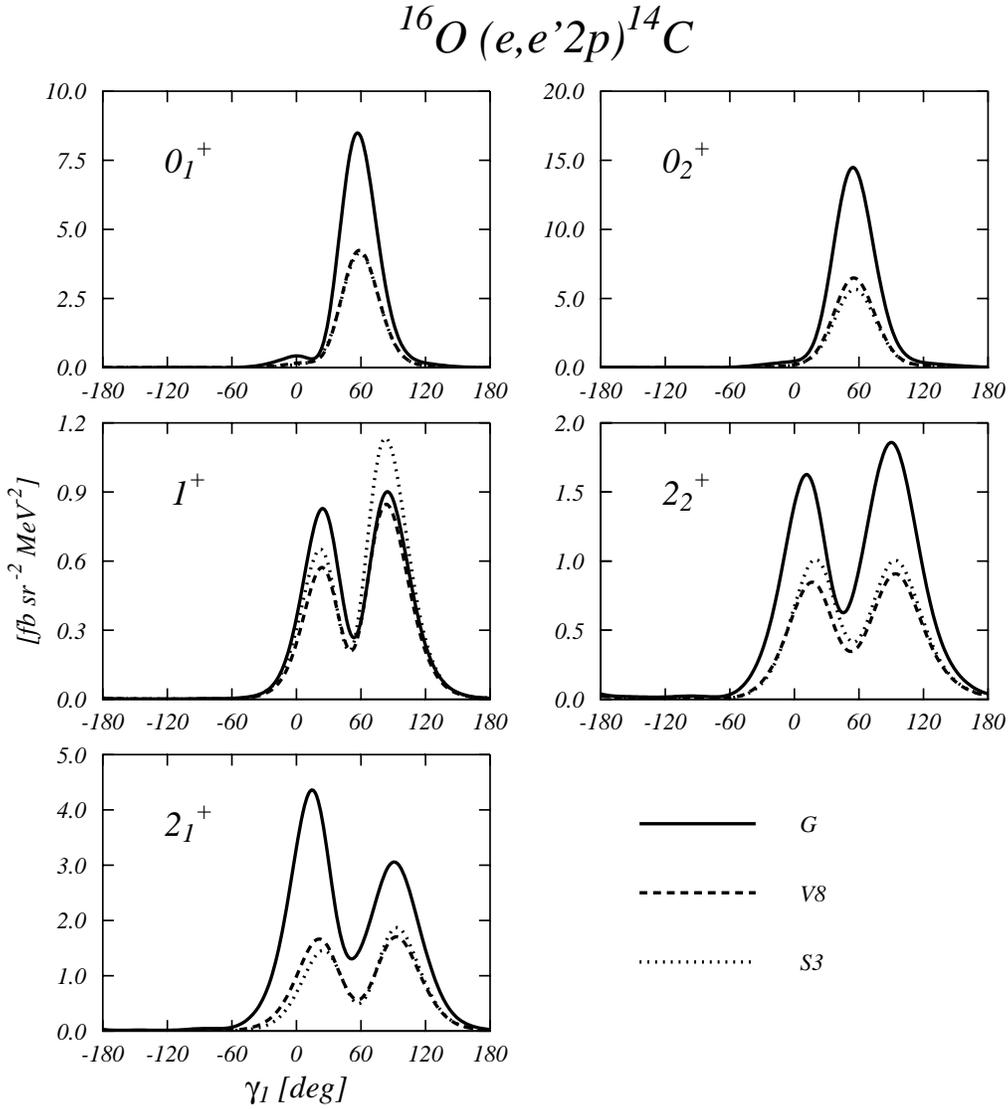}
\caption{\label{fig:s3v8} \small $^{16}$O(e,e'2p)$^{14}$C cross
sections calculated with the correlation functions shown in
Fig. \protect\ref{fig:corr}. The calculations have been performed for
an electron incoming energy $\epsilon$ = 800 MeV, a momentum transfer
$q=400$~MeV/$c$ and a nucleus excitation energy $\omega=100$ MeV. The
second proton was emitted with an energy $\epsilon_2$ = 40 MeV and
within an angle $\gamma_{\ptw}$ = 60$^{\rm o}$. No recoil of the
residual nucleus was considered and coplanar kinematics ($\phi_{\pon}
= \phi_{\ptw} = 0$) was assumed. The angular momenta refer to the
state of the residual nucleus: $(1p_{1/2})^{-2}:0^+_1$,
$(1p_{3/2})^{-2}:0^+_2$, $(1p_{1/2})^{-1}(1p_{3/2})^{-1}:1^+$,
$(1p_{1/2})^{-1}(1p_{3/2})^{-1}:2^+_1$ and $(1p_{3/2})^{-2}:2^+_2$}
\end{center}
\end{figure}

The minimum of the G and S3 interaction is almost the same, while the
V8 interaction has a deeper value. The V8 and S3 correlations
overshoot the asymptotic value of 1 in the region between r=1 and 2
fm. The cross sections obtained with these correlations for $^{16}$O
are shown in Fig. \ref{fig:s3v8}. Apart the result of the 1$^+$ state
dominated by the two-body currents, all the other results have common
trends.  First, one should notice that the use of different
correlations does not change the shape of the angular
distribution. Second, the cross sections obtained with the Gaussian
correlation, are larger than the other ones.

\begin{figure}[htb]
\begin{center}
\includegraphics[height=0.70\textheight]{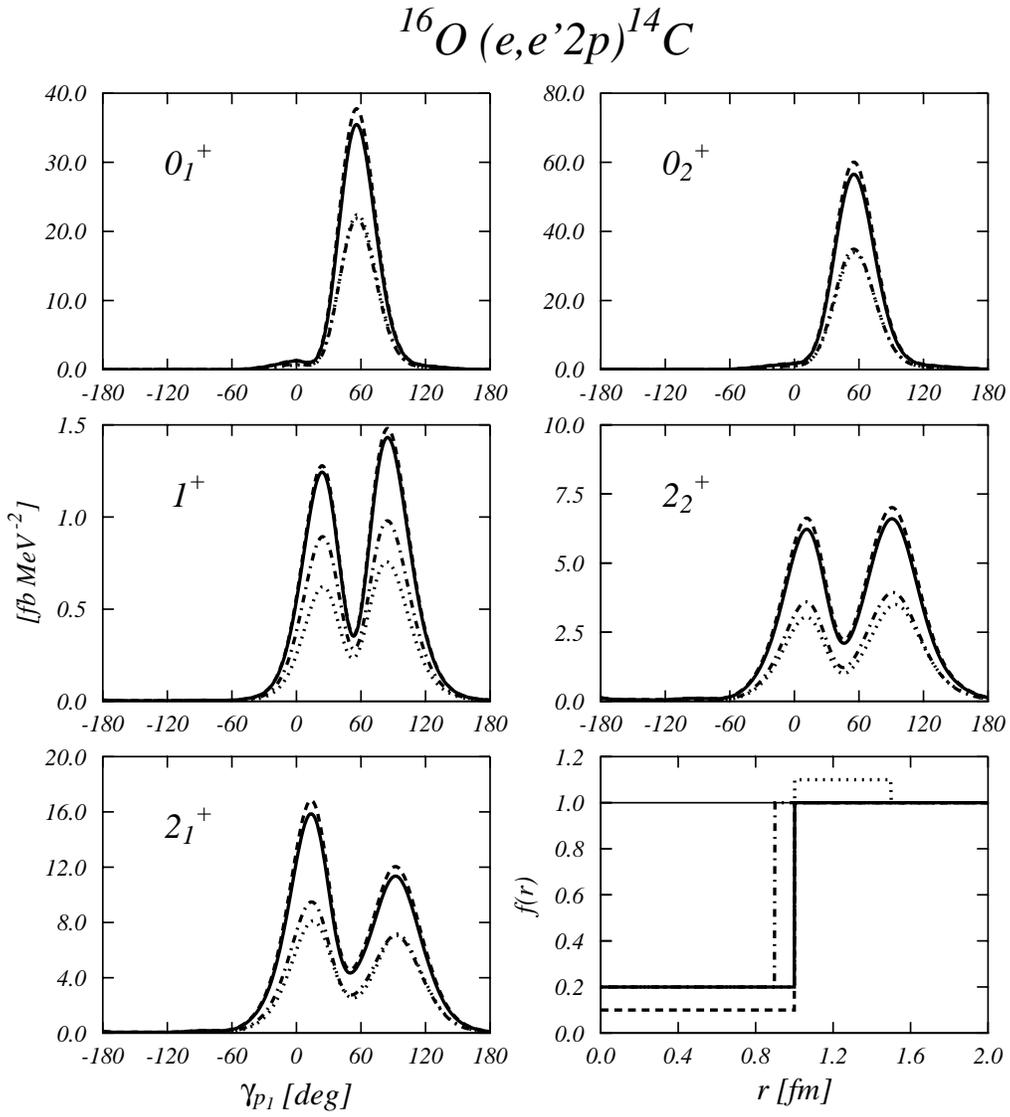}
\caption{\label{fig:box} \small Same as in
Fig. \protect\ref{fig:s3v8} but for the schematic correlations shown
in the lower right panel.}
\end{center}
\end{figure}

To understand this result, we have done a set of calculations with
rather schematic correlations. These correlations are shown in the
lower right panel of Fig. \ref{fig:box}.  The cross sections
calculated with these correlations are shown in the other panels by
lines of the same type. In these calculations the two-body $\Delta$
currents have not been included.

The box correlation indicated by the full line is our reference
correlation. Lowering the minimum (dashed lines) does not produce a
large effect. The insertion of a part which overshoots the asymptotic
value reduces the cross section (dotted lines). An analogous effect is
obtained by reducing the size of the box (dashed-dotted lines).

These results can be understood by remembering that the quantity
entering in the cross section calculation is $h(r)=1 - f^2(r)$ . The
largest is the contribution of $h$ to $\xi_n^1(\bqu)$, the largest
is the cross section. The overshooting of the asymptotic value,
generates a term in $h(r)$ of opposite sign with respect to the rest
of the function, therefore the total contribution to the integral
becomes smaller. The same effect can be obtained by reducing the size
of the box as it is shown by the dashed dotted lines.

\begin{figure}[htb]
\begin{center}
\includegraphics[height=0.70\textheight]{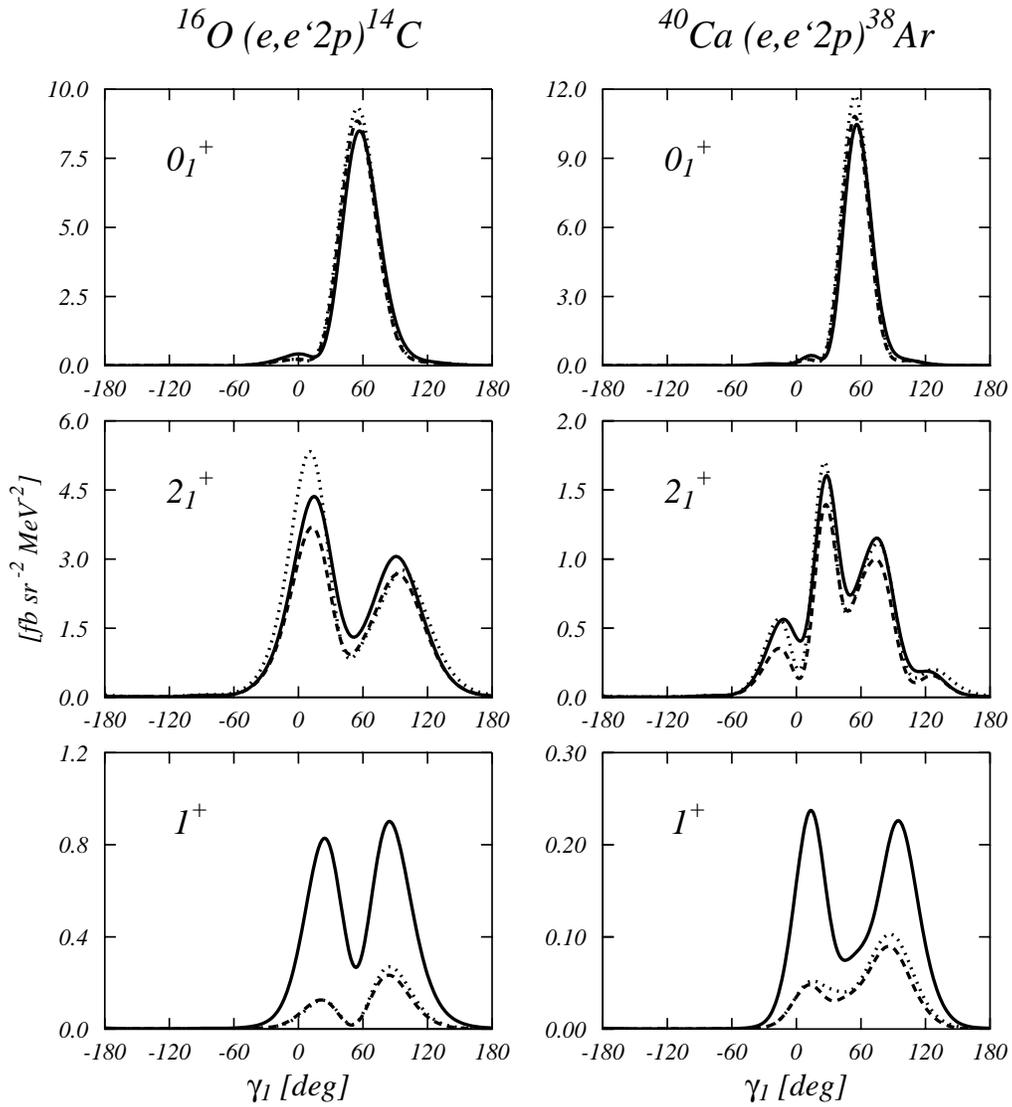}
\caption{\label{fig:2p3pdd} \small $^{16}$O(e,e'2p)$^{14}$C and
$^{40}$Ca(e,e'2p)$^{38}$Ar cross sections calculated with the same
kinematics as in Fig. \protect\ref{fig:s3v8} and with the Gaussian
correlation. Dotted curves include only the 2-point diagrams of
Fig. \ref{fig:diag-2p2h}. Dashed curves include also the 3-point of
the same figure (except the last four which so not contribute to this
process). Solid curves show the results when also the $\Delta$
currents are included.}
\end{center}
\end{figure}

To finish, we show in Fig. \ref{fig:2p3pdd} the effect of each of the
three ingredients of the transition operator: the two- and three-point
diagrams coming from the one-body charge and current operators and the
$\Delta$-isobar current. As we can see, this last dominates the $1^+$
results, while for the $0^+_1$ the terms associated to SRC (in
particular, these corresponding to 2-point diagrams) produce the
larger contributions. In the case of the $2^+_1$, the 3-point
diagrams are also relevant.

\section{Conclusions}

We have presented here results for the (e,e'2p) process in $^{16}$O
and $^{40}$Ca obtained within a model which includes the effect of
scalar SRC through diagrams with only one correlation line.

We have found that qualitative features, such as the shape of the
angular distributions, are not sensitive to the details of the SRC.
On the other hand, slight differences in the correlation function
produce large modifications in the size of the cross sections.

However, the information about the SRC can be obtained only by a
quantitative comparison between theoretical predictions and
experimental data. Unfortunately the quantitative evaluation of the
(e,e'2p) cross sections suffers from the uncertainties inherent to the
required theoretical input. 

These kind of experiments are not the unique tool to extract the
characteristics of the SRC correlations, but another procedure
allowing, together with elastic, inclusive and one-nucleon emission
experiments, to obtain the information we are looking for. We are
trying to describe all these process in a unique and coherent
theoretical framework, thus permitting meaningful analysis of all of
them.

\end{document}